# PERFORMANCE IMPROVEMENT OF MULTIPLE CONNECTIONS IN AODV WITH THE CONCERN OF NODE BANDWIDTH

Preetha K G[1], A Unnikrishnan[2] and K Poulose Jacob[3]

[1]Department of Information Technology, Rajagiri School of Engineering & Technology, Cochin, India
`preetha_kg@rajagiritech.ac.in`
[2]DRDO, Cochin, India
`unnikrishnan_a@live.com`
[3]Department of Computer Science, Cochin University of Science & Technology, Cochin, India
`kpj@cusat.ac.in`

## ABSTRACT

*Mobile Ad-hoc Networks (MANETS) consists of a collection of mobile nodes without having a central coordination. In MANET, node mobility and dynamic topology play an important role in the performance. MANET provide a solution for network connection at anywhere and at any time. The major features of MANET are quick set up, self organization and self maintenance. Routing is a major challenge in MANET due to it's dynamic topology and high mobility. Several routing algorithms have been developed for routing. This paper studies the AODV protocol and how AODV is performed under multiple connections in the network. Several issues have been identified. The bandwidth is recognized as the prominent factor reducing the performance of the network. This paper gives an improvement of normal AODV for simultaneous multiple connections under the consideration of bandwidth of node.*

## 1. INTRODUCTION

Nowadays, the usage of wireless devices like PDA, mobile phones and laptops etc. and its available services are tremendously increased. MANET plays important roles in these areas. MANET is a self organizing network without having a predefined infrastructure. The nodes are highly mobile in MANET and the placement of node is depend on the application and is unpredictable. The nodes and routing are not controlled by any central node or router. Every node is acting as router or source and the control is distributed among nodes.

In spite of this rising interest MANET imposes serious challenges in routing due to unlimited mobility of nodes and dynamic topology. Due to the limited bandwidth path failures are very frequent in nature. This degrades the performance and throughput of the network significantly. Wireless medium is shared by many users so number collisions, contentions and chances of errors are more in MAMET. Routing is a major issue in MANET due to the lack of central coordination [1]. Various routing protocols have been developed by researchers timely. Broadly these protocols are classified into three categories which are proactive, reactive and hybrid [2]. Proactive routing protocols are the extension of the protocols in wired network. This is also called table driven protocols. Each node creates a routing table in advance and exchanges the table with neighbors periodically. Every node updates its routing table accordingly. Up to date information is always available in proactive protocols. This exchanging of routing table induces extra overhead in the network. DSDV is an example of table driven protocol. In reactive or on demand protocol [5,6] the routing table is created and updated only when the node need to send any data. Whenever a node wants to send data to some other node it first establishes the route to that particular node and update its routing table. This reduces the routing overhead. But in this protocol the routing information is not readily available so there is some delay to find the route. AODV, DSR, TORA are the examples of this category. The third classification is hybrid protocol. This combines the advantages of both proactive and reactive protocols. In this the whole network is divided into zones and inside the zone proactive protocol is used and the node wants to communicate to a node in another zone by using reactive protocol. ZRP is an example of hybrid protocol.

Ad-hoc On-demand Distance Vector (AODV) protocol [3] is one of the important on demand routing protocol in MANET. Many improvements on AODV have been going on for the last few years. The main purpose of this paper is to study the connection establishment in AODV and how multiple connections are implemented. This paper also tries to explore the impact of load/ bandwidth of a node in multiple connections. The paper gives an improvement on AODV with the consideration of load of each node in the network. This ensures the maximum possibility of connection establishment and maintenance without any interference.

The rest of this paper is organized as follows. Section 2 gives the idea of connection establishment and connection maintenance in AODV. Section 3 describes how the multiple connections implemented in MANET. The issues and challenges are addressed in section 4. The proposal for improvement is discussed in section 5. Future enhancement is given in section 6 and a conclusion is given in section 7.

## 2. AD-HOC ON-DEMAND DISTANCE VECTOR ROUTING- A BRIEF IDEA

As discussed in section 1 AODV is an on demand routing protocol. In this, the route between source and destination is established only when the node need to send some data. This avoids the periodic propagation of routing table as in proactive protocol. Reactive protocol is better than proactive in terms of routing overhead, packet delivery ratio and energy efficiency. But route information is not available in advance so for establishing route it takes some time. Route establishment and route maintenance are the two major phases in AODV protocol [1]. During the route establishment phase, source node tries to find a route to the destination. Researchers mainly concentrate on the improvement of AODV with single connection. Multiple connection is differ from single connection in terms of the route between the source to number of destinations. In single connection only one route is established and in multiple connection one source node established different path to different destination simultaneously. The objective of this paper is the study of AODV, which supports multiple connections and the importance of bandwidth of a node in multiple connections.

AODV [2] is the extension of DSDV protocol. DSDV is a proactive protocol whereas AODV is reactive. In DSDV each node maintains complete routing table and update this information periodically. But in AODV each node calculates the route on an on demand basis. This is a type of next hop routing protocol. Each node keeps a routing table and it contains next hop to be used to reach destination. On receiving route request packet, each node checks its routing table whether a valid route to the destination exists then only it forwards the packet to the next hop.
Two control packets RREQ and RREP are mainly used for route establishment process. During the route discovery phase the source node flood route request packet (RREQ) to its neighbors. On receiving RREQ packet node check its routing table if it contains the information about destination then it send a Route reply (RREP) packet to the source node otherwise it forward to its neighbors.  This process repeated until it reaches the destination. If there is no reply within certain time interval the source node assumes that there is no route to the destination is available. In this case, after a certain interval of time source node again broadcast the RREQ message for route discovery. Each RREQ packet is having a sequence number attached to it. This is used to distinguish the new RREQ packet from the old one. If the RREQ packet contains the old sequence number reached at any node it discards the packet immediately.  If the node has not seen the sequence number before then it either forward the message or setup a reverse path to the source. The intermediate node updates its routing table with the address of the neighbor from which the first copy of the broadcast packet received. Whenever the RREQ packet reaches at the destination the full reverse path is established and data can be send through this path. The life time of route can be determined by using a timer. Each route is associated with a timer and at the time of route establishment the associated timer starts and the path is deleted when it expires. Then the route discovery process is repeated from the beginning.  This always gives an extra time delay but reduces the number of packet exchange in the network. The bandwidth of radio waves is very limited so the maximum reduction of network overhead is advantageous.

Route discovery process is depicted in Figure 1. The source S connected to destination D through different path. D is respond to S's RREQ message by establishing the reverse path < D, N3, N2, N1, S> by sending send RREP message. Then node S can send the data through the path < S, N1, N2, N3, D>. Thus the path became forward path.  There are many paths exist between S and D. One of them is selected depend on the earliest response time and number of nodes in the path.

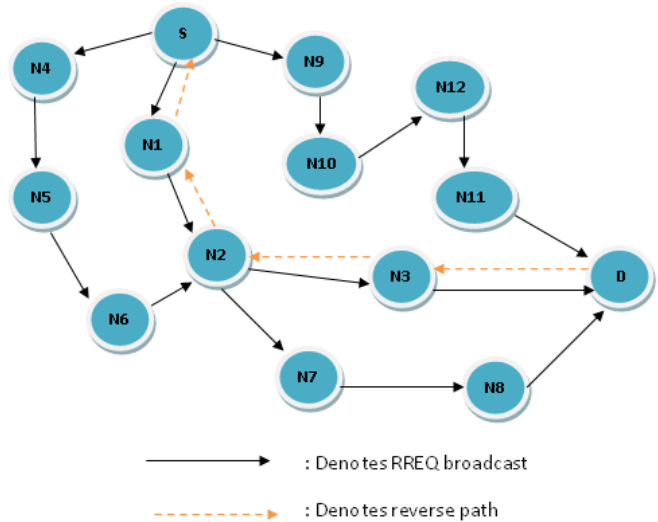

Figure 1: The route discovery process in AODV

After establishing the route form source to destination the data dissemination takes place. The link failures are more frequent in MANET so during the data delivery any node can leave at any time. The link breakage is handled by route maintenance phase of AODV. In order to detect the success of link each node periodically send HELLO packet to its neighbors. A node receives a HELLO packet it update its routing table as the originator of the packet is its neighbor. If HELLO packet reaches safely then the node send back the replay message. The failure of receiving reply for HELLO packet indicates that the neighbor has moved away. So the linked to this neighbor is marked as broken in the routing table. This is reported to all nodes in the network immediately by sending failure message.

The AODV protocol outer perform DSDV in terms of routing overhead [5]. ADOV achieve this by it's on- demand property for route discovery and route maintenance. In spite of this many issues has to be taken care of like time delay for establishing the route and unnecessary flooding of RREQ message for route discovery etc.

## 3. ESTABLISHMENT OF MULTIPLE CONNECTIONS IN AODV

Researchers from all over the world are interested to enhance the performance of AODV from the beginning of this millennium. But most of them consider only single connection from source to destination. Multiple connections are very well suited in real time traffic. In multiple connections one node can communicate with more than one destination at the same time. But this is different from multicasting and multipath routing [16]. In multicasting the same message is transmitted to group of nodes. In multiple connection scenario one node send different messages to different destinations. Multiple connections are used when one node want to communicate to another node at the same time if it wants to share some files to some other node. In this case the same node should connect to two different nodes at the same time.

Theoretically route establishment of both single connection and multiple connections are same [15]. Each connection is considered as individual one. During the link failure, connection establishment algorithm is run on each connection separately. Example of multiple connections is depicted in the figure 2.

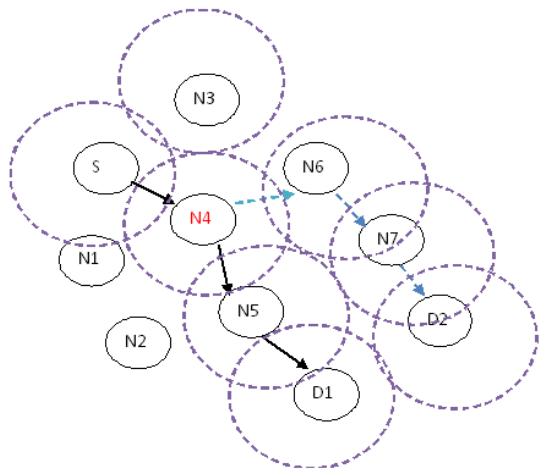

Figure 2. Multiple connections from source S

In the figure 2 source S connect to two destinations D1 and D2 through two different paths <S N4 N5 D1> and <S N4 N6 N7 D2>. Source S can simultaneously communicate to destinations D1 and D2. T he doted circle denotes the communication range. Route establishment algorithm is same as discussed in section 2. Each connection is considered as separate connections and source

S initiate the route discovery procedure by flooding RREQ control packet and find the route to destinations D1 and D2. Route maintenance is little bit difficult in this situation. If any link failures happened then this is reported to the source and then reestablishes the rout. Multiple connections are similar to single connection in performance when the network load is low. The load is increased then several issues to be solved. Possible issues and challenges are discussed in the next session.

## 4. ISSUES IN MULTIPLE CONNECTIONS

Researchers mainly concentrate on MANET with single connections. Lots of improvements have been made for about last decade. Major challenge faced by the researchers in multiple connections is efficient implementation of simultaneous connection establishment [15]. If node failure or link failure happened then the problem became more serious. In single connection the source node concentrate only on the route to a single destination. But in multiple connections more than one route should be taken care of. If a node moves away from the network and this node is the intermediate node of more than one connection then every route should be re-established. Reestablishment also depends on computational power, battery power and time. Simultaneous attempt to establish more than one connection could deteriorate the performance of the network adversely. If more than one connection exists from a node in the network then which connection should establish first is an important question. In order to avoid this crisis we can assign priorities to each process. Real time process has given higher priorities.

Other issue in the multiple connections is the number of connections of a node that can handle. In this paper bandwidth of a node denotes how many routes that passes through the node. If nodes in each route are different from others then it is similar to single connections. In this situation battery power and computational power of the nodes can be utilized in its maximum. If more than one route contains the same node then the power of the node is shared. In figure 2, node N4 is common node for two connections S→D1 and S→D2. This is treated as a serious problem in the multiple connection environments. In this case the capacity of a node is shared for different connections. In some cases, the node energy is not sufficient for all connections passing through it. More than one route shares the same node then the battery power and computational capability is significantly reduced. This affects the overall performance of the network. This paper considers the number of connections that a node can tolerate is also a considering factor of establishing the route. If the bandwidth is ample enough to withstand all the connections then the connection is established. Otherwise search another path.

The above mentioned issues have been the inspiration of developing a new approach to establish the multiple connections in MANET.

## 5. PROPOSAL FOR IMPROVEMENT

Multiple connections are used for simultaneous data transmission from one source to many destinations. In this, simultaneous transmission is considered as separate single connections unlike in multicasting. In multicasting same message is transferred to many destinations. In multiple connection destinations and messages are different. With this, one node can do simultaneous activities within the network at the same time. This always improves the performance in real time situations. In single connections one node initiates one connection at a time. Each activity must wait for the completion of the existing. Even though the performance is improved in multiple connections it has several issues to be solved. It has been observed that the bandwidth of a node is having a vital role during the creation of multiple connections in the

51

network. The novel idea of this proposal is the selection of a node in route establishment by considering its capacity to withstand the connection. In this the capacity is measured in terms of number of connections it already has. In this new scheme, before establish a new route check whether the intermediate node is already participate in any other route. If so calculate its routing capability according to the number of routes it involved and made a check against the maximum limit. Maximum limit is a threshold assigned to each node in the network. Figure 3 represent a typical multiple connection scenario.

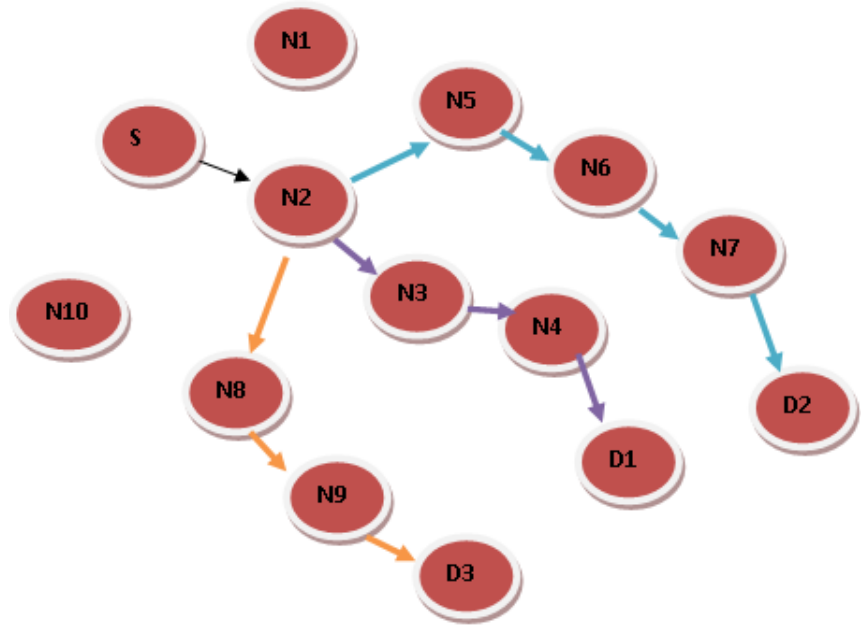

Figure 3. The impact of number of connections of a node

In the above figure, node N2 is a part of 3 connections. Source node S starts route discovery process by sending RREQ packet. Each route is established independently. Node N2 serve as intermediate node in all routes. Consider the situation that there are two routes already existing which is from S to D1 and D2. Node S tries to establish a route to D3. There is only one path to D3 is < N2, N8, N9, D3>. This also contains the node N2. But N2 has already involved in two other connections. If the third connection is permitted then the performance of other two connections are degraded. In order to maintain the performance of the existing connection the third route will not be allowed.

In order to address this problem, before establishing each route the source node always check whether the new connection can be affordable or not. The capability of the node is also a factor of determining the connection establishment. If the intermediate node is capable enough to withstand all the connections including the new one, then the route is established. Otherwise the route should not establish and the route establishment process repeated from the beginning to find another optimal route. In this method each node is assigned a maximum connectivity index. If a route reply comes then source node check against this index of each node in the path before allow to establish a route. If any node in the path already reaches the limit it discards that route and stars new route discovery. To accomplish this method each node adds one more field in the routing table for connection index as < Connection _Index>. The connection index is forward with the HELLO packet then the routing table can be updated accordingly. Set a certain threshold <Route_

limit> denotes the maximum connections of each node beforehand. This indicates the maximum capability of the node that can support the number of connections. The source node establishes the connection only when all the nodes in the route is having the capacity to withstand the new connection. If the connection is established the connection_ index is incremented by one and change is informed to neighbours.

The routing table for node S is given in table 1. Based on figure 3 let us assume that there are two connections from S already established. These are S→D1 and S→D2. According to this the routing table is created.

Table1: Routing table of node S at the time of requesting connection S to D3

| Node | Connection Index |
|------|------------------|
| **S** | **1** |
| **N1** | **0** |
| **N2** | **2** |
| **N3** | **1** |
| **N4** | **1** |
| **N5** | **1** |
| **N6** | **1** |
| **N7** | **1** |
| **N8** | **1** |
| **N9** | **1** |
| **N10** | **0** |
| **D1** | **1** |
| **D2** | **1** |
| **D3** | **0** |

If node S wants to establish a new connection to D3 by using AODV algorithm then it first flood RREQ message in the network. Then the RREP message came back through the reverse path <D3 N9 N8 N2 S>. In normal case the forward path is established from the source to destination and data can send through this route. In the new algorithm after establishing the reverse path the source node check against the threshold value of each node in the route. Here the limit 2 is set for each node. This limit indicates the maximum connection that a node can withstand is 2. The connection <S N2 N8 N9 D3> could not establish because N2 exceeds the maximum limit. In this case S again starts to find another route in the network.

If a node involves more than one connection from the same source there is a possibility of performance degradation. This can be avoided by considering the capacity of a node which can tolerate the number of connection without disturbing the existing operations. In this approach the new route is established without interrupting the existing connections. The proposal is slightly

International Journal on AdHoc Networking Systems (IJANS) Vol. 2, No. 3, July 2012

costly in terms of routing overhead by sending the updated route count, but this overhead is tolerable if the enhanced performance is considered.

## 6. FUTURE ENHANCEMENTS

In this paper multiple connections in AODV is improved with the consideration of number of connection in which a node can handle. In this approach assign the maximum number of connections that can afford to each node. In future, previous experience of the node can also be incorporated with the number of connections. If more than one node competes for a particular destination then its previous experience in the network and packet delivery ratio can also consider for selection. This can also avoid the unnecessary flooding of RREQ packet. If a node willing to forward RREQ packet and calculate the probability of win is less than certain threshold then it will not forward the message otherwise it will forward to its neighbours.

## 7. CONCLUSIONS

This paper studied the route establishment phase in AODV protocol in detail and how the same algorithm can be applied to multiple connection scenarios. The paper listed various issues in multiple connections and node energy is identified as the major concern while establishing simultaneous route to multiple destinations. If the node bandwidth is consider as a metric for establishing the connection then the performance of the network is significantly improved. The note to be kept is that the usage of this algorithm should be encouraged in all sorts of communications, so that it results in more beneficial wireless networks at all levels. The usage of this algorithm will lead to extend the connectivity in wireless networks thus creating an environment of communication in an easy way.

## Authors

Ms. Preetha K G has completed her B Tech and M Tech Degree in Computer Science from Calicut University and Dr. MGR University respectively. She is associated with Rajagiri School of Engineering & Technology as an Assistant Professor in the department of Information Technology. She has around twelve years of academic experience. Currently she is a research scholar in Cochin University of science and Technology. Her Research interests includes Mobile Computing, Wireless Networks, Ad-hoc Networks Etc.

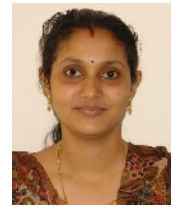

Dr. A Unnikrishnan- Graduated from REC (Calicut), India in Electrical Engineering (1975), completed his M.Tech from IIT, Kanpur in Electrical Engineering (1978) and Ph.D from IISc, Bangalore in "Image Data Structures" (1988). Presently, he is the Associate Director in Naval Physical and Oceanographical Laboratory, Kochi which is a premiere laboratory of defence research and development organisation. His field of interests includes Sonar Signal Processing, Image Processing and Soft Computing. He has authored about fifty National and International Journal and Conference Papers. He is a Fellow of IETE & IEI, India.

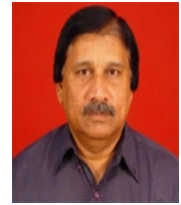

Dr. K.Poulose Jacob, Professor of Computer Science at Cochin University of Science and Technology since 1994, is currently Director of the School of Computer Science Studies. . He was Chairman of two Boards of Studies earlier. He has been the Dean of the Faculty of Engineering and is presently Chairman, Board of Studies in Computer Science. He is a member of Academic Council and has been in the University Senate for 10 years. He has presented research papers in several International Conferences in Europe, USA, UK, Australia and other countries. He has delivered invited talks at several national and international events.

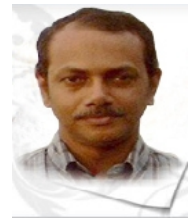

He has served as a Member of the Standing Committee of the UGC on Computer Education & Development. He is the Zonal Coordinator of the DOEACC Society under the Ministry of Information Technology, Government of India. He serves as a member of the AICTE expert panel for accreditation and approval. Dr. K.Poulose Jacob is a Professional member of the ACM (Association for Computing Machinery) and a Life Member of the Computer Society of India.